\documentclass[11pt,a4paper]{article}

\usepackage[T1]{fontenc}
\usepackage[utf8]{inputenc}
\IfFileExists{lmodern.sty}{\usepackage{lmodern}\usepackage{microtype}}{}
\usepackage{amsmath,amssymb,amsthm}
\usepackage{mathtools}
\usepackage[margin=2.5cm]{geometry}
\usepackage{booktabs}
\usepackage{tikz}
\usetikzlibrary{arrows.meta,positioning,bending}
\usepackage{enumitem}
\usepackage[round]{natbib}
\usepackage[colorlinks=true,linkcolor=blue!55!black,citecolor=blue!55!black,urlcolor=blue!55!black]{hyperref}
\theoremstyle{plain}
\newtheorem{theorem}{Theorem}[section]
\newtheorem{proposition}[theorem]{Proposition}
\newtheorem{lemma}[theorem]{Lemma}
\theoremstyle{definition}
\newtheorem{definition}[theorem]{Definition}
\newtheorem{example}[theorem]{Example}

\newcommand{\V}{\mathbf{V}}
\newcommand{\Vz}{\mathbf{V}_0}
\newcommand{\X}{\mathbf{X}}
\newcommand{\Y}{\mathbf{Y}}
\newcommand{\Es}{\mathbf{S}}
\newcommand{\El}{\mathbf{S}_{L}}
\newcommand{\Ew}{\mathbf{W}}
\newcommand{\Ub}{\mathbf{U}}
\newcommand{\Kb}{\mathbf{K}}
\newcommand{\Sp}{\mathbf{S}^{+}}
\newcommand{\Mb}{\mathbf{M}}
\newcommand{\Ab}{\mathbf{A}}
\newcommand{\Bb}{\mathbf{B}}
\newcommand{\Rb}{\mathbf{R}}
\newcommand{\Eb}{\mathbf{E}}
\newcommand{\Db}{\mathbf{D}}
\newcommand{\An}{\mathit{An}}
\newcommand{\Pa}{\mathit{Pa}}
\newcommand{\Ccomp}{\mathrm{C}}
\newcommand{\pre}[1]{\V^{(#1)}}

\tikzset{
  dirT/.style={-{Stealth[length=2.6mm]},line width=1.1pt},
  vtx/.style={circle,draw,inner sep=1.2pt,minimum size=5.5mm,font=\footnotesize},
  dir/.style={-{Stealth[length=2mm]},thin},
  bid/.style={{Stealth[length=2mm]}-{Stealth[length=2mm]},dashed,thin,gray},
  call/.style={draw,rounded corners=2pt,inner sep=3pt,font=\scriptsize,align=center}
}

\title{On the trajectories of the ID algorithm}

\author{%
  Ivano Lodato\textsuperscript{1},\quad
  Simran Hiranandani\textsuperscript{1},\quad
  Aditya Varna Iyer\textsuperscript{1,2}\\[0.8em]
  {\small\textsuperscript{1}Allos AI Limited, Hatton Garden, London, EC1 N8LE, UK}\\
  {\small\textsuperscript{2}Clarendon Laboratory, Department of Physics, University of Oxford, Oxford OX1 3PU, UK}\\[0.4em]
  {\small\texttt{ivano.lodato@allos.ai}, \texttt{simran.h@allos.ai},
   \texttt{aditya.iyer@allos.ai}}%
}
\date{}

\begin{document}
\maketitle

\begin{abstract}
\noindent The ID algorithm decides whether an interventional distribution can be computed from
observational data and, when it can, returns a formula in which products, ratios and summations of
conditional probabilities are nested inside one another. We study these formulas by following the
trajectories of the calls of the algorithm, each of which ends with a tail (an alternation of lines 7
and 2) and an exit, and we derive two main results. First, line~4 and the exit lines fix the single
outer summation of the output, while the tail forms everything inside it, both the inner summations
and the ratios. Second, a ratio keeps a summation in its denominator only when a repetition of the
tail is entangled, and a test of ancestry in the original graph decides, before any formula is
computed, whether a topological order can be chosen so as to avoid it. We also derive a bound on the
nesting depth of the output, and show a simple ``dimensional'' count that every output has to satisfy.
Together, these results constrain the form of an ID output and guide its implementation and
simplification.
\end{abstract}

\section{Introduction}
\label{sec:intro}

The effect of an action, a decision or a policy is most reliably measured by an experiment, but
experiments are often costly, unethical or infeasible, and one is then led to ask whether the same
effect can be inferred from data that have merely been observed. The question acquires a precise form
when the causal assumptions are encoded in a diagram $G$, whose directed edges represent direct causal
influence and whose bidirected edges represent latent common causes. The quantity of interest is the
effect of an intervention, that is, the distribution $P(\Y\mid\hat{\X})$ that a set of outcomes
$\Y$ would have if the variables $\X$ were \emph{set} by an external action rather than observed,
and this effect is \emph{identifiable} if it can be computed from the observational distribution
$P(\V)$ and $G$ alone \citep{pearl1995}. The do-calculus \citep{pearl1995} is complete for this task
\citep{shpitserpearl2006a,huangvaltorta2006}, but it does not prescribe the order in which its rules
should be applied, and the decisive step towards an algorithm was the observation of
\citet{tianpearl2002,tianpearl2003} that $P(\V)$ factorises over the \emph{c-components} of $G$, the
maximal sets of vertices connected by bidirected paths. The ID algorithm of
\citet{shpitserpearl2006a,shpitserpearl2008} turns this factorisation into a recursive procedure of
seven lines, which decides identifiability in polynomial time (Section~\ref{sec:traj}) and, when the
effect is identifiable, returns a formula for it.\footnote{The results of this paper apply, mutatis mutandis, to the conditional queries solved
by the algorithm IDC \citep{shpitserpearl2006b}, because the output of IDC is either an ID output or
the ratio of an ID output to its sum over the outcomes, in both cases for an enlarged set of
interventions.} The correctness of this procedure is well
established, whereas much less is known about the formula that it returns, whose structure has so far
been examined only in order to simplify it, and this formula is the subject of the present paper.

The formula that ID returns is built from (possibly nested) products, ratios and summations of
conditional distributions $P(V\mid C)$, which we call \emph{terms}. A familiar instance is
the front-door graph
\begin{center}
\begin{tikzpicture}[x=1cm,y=1cm]
\node[vtx] (x) at (0,0) {$X$};
\node[vtx] (m) at (1.9,0.95) {$Z$};
\node[vtx] (y) at (3.8,0) {$Y$};
\draw[dir] (x) -- (m);
\draw[dir] (m) -- (y);
\draw[bid] (x) -- (y);
\end{tikzpicture}
\end{center}
for which ID returns
\begin{equation}\label{eq:frontdoor}
  P(Y\mid\hat{X})=\sum_Z\Bigl(P(Z\mid X)\ \sum_X P(X)\,P(Y\mid X,Z)\Bigr),
\end{equation}
a formula whose two summations have different origins. The outer one, over $Z$, is written by line~4
of the algorithm and marginalises the variables that are neither intervened on nor queried. The inner
one, over $X$, arises at a later stage of the execution, where line~7, having isolated the c-component
$\{X,Y\}$ of $Y$, rebuilds a distribution on it, and the line~2 that follows marginalises $X$, which
is not an ancestor of $Y$ within that component. We call such an alternation of lines 7 and 2 a
\emph{tail}, and each pair of lines a \emph{repetition} of it. In larger graphs the tail has several
repetitions, each of which, after the first, takes conditionals of a distribution that already carries
a summation, and this is how ratios of sums come to appear inside further sums. Here and throughout, every variable is written with a capital letter. The
letter $X$ thus occurs in \eqref{eq:frontdoor} in two roles, as the intervened variable in
$P(Z\mid X)$, where it stands for the value set by the intervention, and as a summation variable
inside the brackets.

Any implementation of the ID algorithm, such as the R package \texttt{causaleffect}
\citep{tikkakarvanen2017jss} or the Python library of the same name \citep{pedemonte2021}, has to
store such formulas and, as far as possible, to simplify them. The ID algorithm writes
a summation wherever a marginalisation is called for and a ratio wherever a conditional of a summed distribution is called
for, whether or not the summation can be carried out or the ratio cancelled. Storing such formulas
requires a class of expressions, and the available choices differ. For instance,
\citet{tikkakarvanen2017} nest products and summations of terms into \emph{expressions} and treat a
quotient of two expressions separately, as a last operation, whereas \citet{pedemonte2021} use a
recursive probability object, which may be summed over a set of variables and may carry a divisor of
the same kind. Simplifying those expressions means carrying out summations
\citep{tikkakarvanen2017,companion}, cancelling factors within a single ratio, or cancelling factors
between different ratios. These simplifications are applied to the finished formula or, at best, step
by step while it is assembled, and in either case without regard to the way in which the algorithm
produces it.

Here we study the formula at its source, by following the trajectories of the algorithm, that is, the
sequences of lines through which it arrives at the formula, and we determine where the summations and
the ratios of an output are created, which terms they involve and how deeply they can be nested. The
trajectories, and the sets of vertices on which each of their lines acts, are obtained from the graph and the query by graph operations alone, so that this information is
available before any formula is written; it constrains the expressions that an implementation has to
represent and indicates where each kind of simplification can act. Section~\ref{sec:traj} determines
the trajectories that are possible and where the summations of the output are created. Section~\ref{sec:nest} studies the ratios, which arise when the rebuild-and-marginalise step
of the tail is repeated, and decides on the graph when they can be avoided. Section~\ref{sec:simp} shows what
simplification can do with the ratios that remain, and Section~\ref{sec:depth} bounds the nesting
depth of the output. All the proofs are collected in Appendix~\ref{app:proofs}.

\section{The trajectories of ID}
\label{sec:traj}

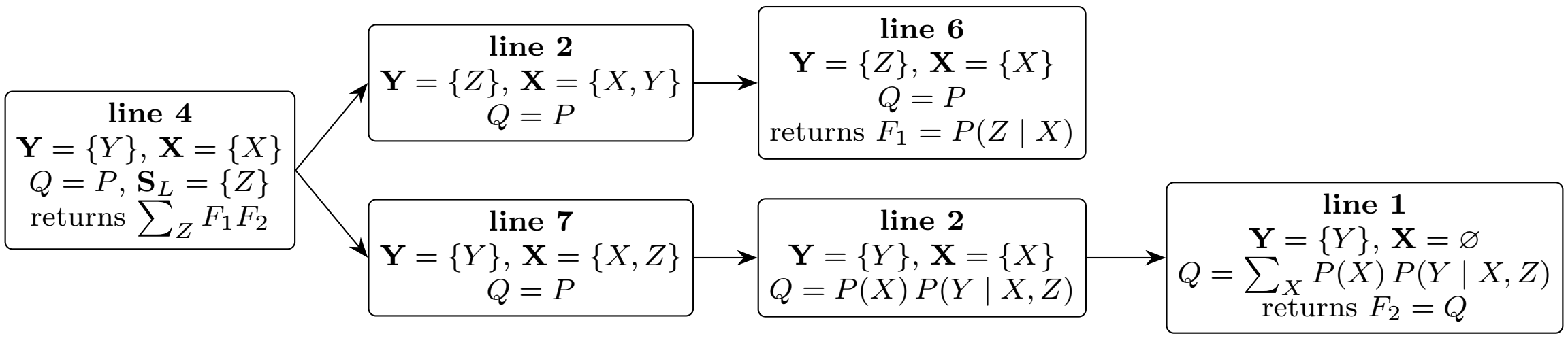
\begin{figure}[!ht]\centering
\begin{tikzpicture}[x=1cm,y=1cm]
\node[call] (r)  at (0,0)      {\textbf{line 4}\\$\Y=\{Y\}$, $\X=\{X\}$\\$Q=P$, $\El=\{Z\}$\\returns $\sum_{Z}F_1F_2$};
\node[call] (a1) at (3.7,0.85) {\textbf{line 2}\\$\Y=\{Z\}$, $\X=\{X,Y\}$\\$Q=P$};
\node[call] (a2) at (7.5,0.85) {\textbf{line 6}\\$\Y=\{Z\}$, $\X=\{X\}$\\$Q=P$\\returns $F_1=P(Z\mid X)$};
\node[call] (b1) at (3.7,-0.85) {\textbf{line 7}\\$\Y=\{Y\}$, $\X=\{X,Z\}$\\$Q=P$};
\node[call] (b2) at (7.5,-0.85) {\textbf{line 2}\\$\Y=\{Y\}$, $\X=\{X\}$\\$Q=P(X)\,P(Y\mid X,Z)$};
\node[call] (b3) at (11.8,-0.85) {\textbf{line 1}\\$\Y=\{Y\}$, $\X=\varnothing$\\$Q=\sum_XP(X)\,P(Y\mid X,Z)$\\returns $F_2=Q$};
\draw[dir] (r.east) -- (a1.west);
\draw[dir] (r.east) -- (b1.west);
\draw[dir] (a1) -- (a2);
\draw[dir] (b1) -- (b2);
\draw[dir] (b2) -- (b3);
\end{tikzpicture}
\caption{The recursion tree of ID for the front-door query, with the line executed by each call and
its arguments. The root, the only branching point, returns the product of the formulas $F_1$ and
$F_2$ returned by its two branches, summed over its leftover set $\El=\{Z\}$.}
\label{fig:tree}
\end{figure}

A \emph{term} is a factor $P(V\mid C)$ with a single variable $V$, its \emph{head}, and a set
$C\not\ni V$, its \emph{conditioning set}. A \emph{call} is an execution of
$\mathrm{ID}(\Y,\X,Q,G)$, with $G=(\V,E)$, in which $\Y$ is the outcome set, $\X$ the intervention
set and $Q$ the \emph{distribution argument}, and it returns a formula built from terms unless it
fails. Each call executes exactly one of the seven lines of the algorithm, namely the first whose test
succeeds. Lines 2, 3 and 7 make one further call, the \emph{child} of the call that executes them,
line~4 makes $k\ge2$ of them, and lines 1 and 6 make none, while line~5 reports failure. The calls
therefore follow one another along a single path or, when line~4 is executed, along the branches of a
tree, the \emph{recursion tree}. A \emph{trajectory} is the sequence of the lines executed along a
path from the root to a leaf. Figure~\ref{fig:tree} shows the recursion tree of the
front-door query, whose trajectories are $4,2,6$ and $4,7,2,1$. Throughout, the query is assumed
identifiable, so that line~5 is never executed, and interventional, so that $\X\neq\varnothing$ at the
root.

We write $P$ for the joint observational distribution of the vertices of the graph at hand. Until a
line~6 or a line~7 is executed, a summation only reduces this joint distribution to the joint
distribution of fewer vertices, which we still denote by $P$ and for which no explicit summation is
written, as in the upper branch of Figure~\ref{fig:tree}. These two lines are the first to break $P$
into terms, and line~7 passes the product of these terms on as the new distribution argument. The
line~2 that follows places on this product a summation that is not a marginalisation of $P$, so that
the distribution argument differs from $P$ only in the tail.

The terms are formed with respect to a topological order $\pi$ of the vertex set $\Vz$ of the query,
which is fixed once and for all, and $u<v$ means that $u$ precedes $v$ in $\pi$. If $V_1<\dots<V_n$ are the vertices of the vertex set $\V$ of a call, we write
$\pre{i}=\{V_1,\dots,V_i\}$, with $\pre{0}=\varnothing$, $T_i$ for the c-component of $V_i$ in the
subgraph of $G$ induced on $\pre{i}$, and
\[
  C_i=\Pa(T_i)\setminus\{V_i\},
\]
where $\Pa(T_i)$ consists of $T_i$ and of the parents in $G$ of its members. Lines 6 and 7 form the
\emph{Tian factors} $Q(V_i\mid C_i)$ of the distribution argument \citep[Lemmas~1 and~4]{tianpearl2003}.
Every member of $C_i$ precedes $V_i$, so that a Tian factor involves only variables up to its head.
When $Q=P$, the Tian factor is the term $P(V_i\mid C_i)$ (whose conditioning set is minimal under
faithfulness; \citealp{tianpearl2003,companion}). In the front-door graph with the order $X<Z<Y$, for
instance, the c-component of $Y$ is $T_3=\{X,Y\}$, the only parent that it adds is $Z$, and
$C_3=\{X,Z\}$, which gives the term $P(Y\mid X,Z)$ of \eqref{eq:frontdoor}. In the tail, where $Q$
is no longer equivalent to $P$, the Tian factors have to be computed from $Q$, and
Section~\ref{sec:nest} is devoted to this computation.

Two sets derived from the arguments of a call govern the analysis, the \emph{live set} $\Es$ and the
\emph{leftover set} $\El$,
\[
  \Es=\V\setminus\X,\qquad \El=\Es\setminus\Y,
\]
which consist of the vertices that are not intervened on and, among these, of those that are not
outcomes; at the root of Figure~\ref{fig:tree}, $\Es=\{Z,Y\}$ and $\El=\{Z\}$. We write $G[\Es]$ for
the subgraph of $G$ induced on $\Es$ and $\Ccomp(G)$ for the set of the c-components of a graph $G$.
When the test of line~4 fails, $G[\Es]$ has a single c-component, which is the set denoted $\Es$ in
lines 5--7 of the pseudocode, and at line~7 $\Sp$ denotes the c-component of $G$ that strictly
contains it. The table below lists, for each line, its test and the formula returned or the
arguments $\V'$, $\Y'$, $\X'$, $Q'$ of its child.

\begin{center}
\renewcommand{\arraystretch}{1.15}
\begin{tabular}{@{}clcccc@{}}
\toprule
line & test & $\V'$ & $\Y'$ & $\X'$ & $Q'$ \\
\midrule
1 & $\X=\varnothing$
  & \multicolumn{4}{c}{return $\sum\nolimits_{\El}Q$} \\
2 & $\V\neq\An(\Y)_G$
  & $\An(\Y)_G$ & $\Y$ & $\X\cap\V'$ & $\sum\nolimits_{\V\setminus\V'}Q$ \\
3 & $\Ew:=\Es\setminus\An(\Y)_{G_{\bar{\X}}}\neq\varnothing$
  & $\V$ & $\Y$ & $\X\cup\Ew$ & $Q$ \\
4 & $\Ccomp(G[\Es])=\{\Es_1,\dots,\Es_k\}$, $k>1$
  & \multicolumn{4}{c}{return
    $\sum\nolimits_{\El}\prod\nolimits_{i=1}^{k}
     \mathrm{ID}\bigl(\Es_i,\ \V\setminus\Es_i,\ Q,\ G\bigr)$} \\
5 & $\Ccomp(G)=\{\V\}$ & \multicolumn{4}{c}{\textsc{fail}} \\
6 & $\Es\in\Ccomp(G)$
  & \multicolumn{4}{c}{return $\sum\nolimits_{\El}\prod\nolimits_{V_i\in\Es}Q(V_i\mid C_i)$} \\
7 & $\Es\subsetneq\Sp\in\Ccomp(G)$
  & $\Sp$ & $\Y$ & $\X\cap\Sp$ & $\prod\nolimits_{V_i\in\Sp}Q(V_i\mid C_i)$ \\
\bottomrule
\end{tabular}
\end{center}

Two features of the table deserve a remark. First, a set $C_i$ formed at line~7 may contain vertices
outside $\Sp$, which in the product passed on by line~7 are no longer variables but \emph{fixed
values}, as $Z$ is in Figure~\ref{fig:tree}; like every variable, they keep their capital letter.
Second, summations enter the output in two ways: lines 1, 4 and 6 return a summation over the leftover
set of their own call (at line~1 this set is $\V\setminus\Y$, because $\X$ is empty), and line~2
places one inside the distribution argument.

The course of a trajectory is controlled by two properties of the live set $\Es$ of a call:
\begin{description}[nosep]
  \item[(A)] every $v\in\Es$ has a directed path to $\Y$ in $G$ with all its vertices in $\Es$, where
    a path may have length zero, so that the vertices of $\Y$ always meet the requirement;
  \item[(B)] $G[\Es]$ has exactly one c-component, that is, $\Ccomp(G[\Es])$ has a single element.
\end{description}
We also say that a call is \emph{saturated} when its leftover set is empty, that is, when $\Es=\Y$, in
which case (A) holds through paths of length zero. At the root of Figure~\ref{fig:tree}, (A) holds
because of the edge $Z\to Y$, whereas (B) fails, since $Z$ and $Y$ are not joined by a bidirected
edge, and the two children of the root are saturated. Conditions (A) and (B) are identified with the
outcomes of the tests of lines 3 and 4, as the following lemma shows.

\begin{lemma}\label{lem:AB}
Condition (A) holds if and only if the test of line~3 fails, and (B) holds if and only if the test of
line~4 fails, so that every call reaching line~5 or beyond satisfies both.
\end{lemma}

Lemma~\ref{lem:AB} describes a single call, whereas to follow a trajectory we also need to know what
becomes of the two conditions when a call hands over to its child or children, which is the content of
the next lemma. As it turns out, once both conditions hold they hold for the rest of the trajectory,
which can then execute only lines 2 and 7 until it exits; the trajectory remains, so to speak,
trapped, and this gives the lemma its name.

\begin{lemma}[Trap Lemma]\label{lem:trap}
Line~2 under (A) and line~7 leave the live set and the graph induced on it unchanged, and therefore
preserve (A), (B) and saturation. Each child of a line~4 is saturated and satisfies (A) and (B),
while line~3 gives its child (A) but not necessarily (B).
\end{lemma}

Together, the two lemmas restrict the order in which the lines can follow one another, and the next
lemma collects these restrictions; its item~\ref{it:7} is the only one that uses identifiability.

\begin{lemma}[Control flow]\label{lem:flow}
Along every trajectory
\begin{enumerate}[label=(\roman*),nosep]
  \item\label{it:22} line~2 is never immediately followed by line~2;
  \item\label{it:1} line~1 is immediately preceded by line~2;
  \item\label{it:3} condition (A) persists once established, and line~3 is executed at most once, at
    step 1 or 2;
  \item\label{it:7} the call immediately following a line~7 executes line~2.
\end{enumerate}
Moreover, line~4 is executed at most once in the entire recursion tree.
\end{lemma}

To see why item~\ref{it:7} holds, note that the child of a line~7 inherits (A) and (B), and that the
test of line~5 succeeds there, because the vertex set of the child is a single c-component of its
graph. Since the query is identifiable, the child has to be stopped by an earlier line, and line~2 is
the only one that it can execute. The three lemmas, which are proved in Appendix~\ref{app:proofs},
leave little freedom to a trajectory, and Theorem~\ref{thm:T} below shows that it consists of the
four phases of Figure~\ref{fig:phases}.

\begin{theorem}[Shape of a trajectory]\label{thm:T}
Every trajectory consists of four consecutive phases: a \emph{clean-up}, one of $\varnothing$, $2$,
$3$, $2\,3$; a \emph{branching}, which is either empty or consists of line~4 followed, separately in
each of its branches, by at most one line~2; a \emph{tail}, in which the pair $7\,2$ is repeated
$m\ge0$ times, each occurrence of the pair being a \emph{repetition}; and an \emph{exit}, line~6 or
line~1, where line~1 is immediately preceded by a line~2 and neither exit is immediately preceded by
line~7. If $\X_0$ is the intervention set of the call at
which the tail begins, then $m\le\lfloor|\X_0|/2\rfloor$.
\end{theorem}

\begin{figure}[!ht]\centering
\begin{tikzpicture}[phase/.style={draw,rounded corners=2pt,minimum height=11mm,inner sep=4pt,
  font=\footnotesize,align=center},note/.style={font=\scriptsize\itshape,align=center,text width=36mm}]
\node[phase] (c) {\textbf{clean-up}\\$\varnothing$,\ \ $2$,\ \ $3$\ \ or\ \ $2\,3$};
\node[phase,right=6mm of c] (b) {\textbf{branching}\\$\varnothing$,\ \ or $4$ and then\\$\varnothing$ or $2$ in each child};
\node[phase,right=6mm of b] (t) {\textbf{tail}\\$7\,2$ repeated $m\ge0$ times};
\node[phase,right=6mm of t] (e) {\textbf{exit}\\$6$\ \ or\ \ $1$};
\draw[dir] (c) -- (b); \draw[dir] (b) -- (t); \draw[dir] (t) -- (e);
\node[note,below=1mm of c] {discards vertices irrelevant to $\Y$};
\node[note,below=1mm of b] {splits $\Es$, sums over $\El$};
\node[note,below=1mm of t] {rebuilds, then marginalises};
\node[note,below=1mm of e] {returns};
\end{tikzpicture}
\caption{The four phases of a trajectory for an identifiable interventional query
(Theorem~\ref{thm:T}).}
\label{fig:phases}
\end{figure}
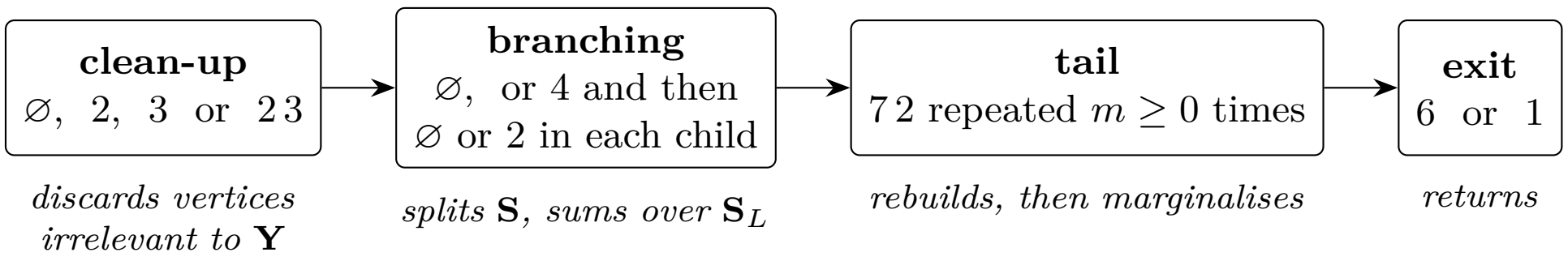

Consider again Figure~\ref{fig:tree}: the upper trajectory has the branching $4\,2$, an empty tail and
the exit $6$, and the lower one has the branching $4$, a tail with $m=1$ and the exit $1$.
Theorem~\ref{thm:T} also bounds the total number of calls in the trajectories of a query, because it
bounds each phase separately. The clean-up and line~4 account for at most three calls, and they are followed by $k$
branches ($k=1$ if line~4 is not executed), each of which consists of at most one line~2, of the $2m$
calls of its tail and of one exit. Since $2m\le|\X_0|<|\Vz|$ (at least one vertex is an outcome),
the recursion tree has at most $3+k(|\Vz|+1)$ calls, with $k<|\Vz|$. Each call computes a bounded
number of ancestral sets and of decompositions into c-components, so that ID decides identifiability
in polynomial time, as \citet{tian2004} observes for his identification algorithm.

Finally, Theorem~\ref{thm:T} locates the two kinds of summation of an output, those over a leftover
set $\El$ and those that a line~2 places inside the distribution argument. The former are written at
line~4 and at the exits, but if line~4 is executed the
exits of its branches sum over the empty set, because every call after the branching is saturated
(Lemma~\ref{lem:trap}). The output thus carries exactly one such summation, written by line~4 if it
is executed and by the single exit otherwise, and in both cases it runs over the leftover set of the
call at which the clean-up ends, since neither the live set nor the outcome set changes along the
tail. The latter remain in the output only when the line~2 belongs to a tail, where it acts on the
product built by the preceding line~7. The following proposition records these observations, and expresses the set of the
outer summation through the original graph alone.

\begin{proposition}\label{prop:onesum}
Every ID output has the form $\sum_{\El}\Phi$, with
$\El=\An(\Y)_{G_0[\Vz\setminus\X]}\setminus\Y$, where $\Y$, $\X$ and $G_0$ are the outcome set, the
intervention set and the graph of the query. The summation over $\El$ is written by line~4, if line~4
is executed, in which case $\Phi$ is the product of the formulas returned by its branches, and by the
exit of the only trajectory otherwise. Every summation inside $\Phi$ is created by a line~2 of a tail
and acts on the product built by the line~7 that precedes it.
\end{proposition}

In the front-door formula \eqref{eq:frontdoor}, line~4 is executed, the root sums over $\El=\{Z\}$,
and the summation over $X$ is created by the line~2 of the lower trajectory of
Figure~\ref{fig:tree}. When line~4 is not executed, the same set is summed by either of the exit
lines 1 and 6. If the bidirected edge of the front-door graph joins $Z$ and $Y$ instead of $X$ and $Y$, for instance, the
query $P(Y\mid\hat{X})$ is answered by line~6 alone, which returns
$\sum_ZP(Z\mid X)\,P(Y\mid X,Z)$, again with $\El=\{Z\}$. In Example~\ref{ex:tail} below line~4 is
not executed either, but the exit is saturated and all three summations arise in the tail.

\section{Ratios along the tail}
\label{sec:nest}

As we have seen in Section~\ref{sec:traj}, and in Theorem~\ref{thm:T} in particular, the topological
order $\pi$ plays no role in the trajectories. Every test, and every set passed to a
child, is computed from $\V$, $\Y$, $\X$ and $G$ alone, so that the recursion tree, its trajectories
and the sets on which each line acts are obtained from the graph and the query by graph operations,
before any formula is written. The order enters only through the Tian factors formed at lines 6 and
7, and this section shows how these factors are written along the tail, when they are ratios, and
when the order can prevent it.

\subsection{The Tian factors of a summed product}

The first call of a trajectory that executes line~6 or line~7 has $Q=P$, and its Tian factors are the
terms $P(V_i\mid C_i)$. If this call executes line~7, the line~2 that follows sums out some vertices,
and from then on the distribution argument is a summed product, which is not a marginal of $P$. Its
Tian factors are given by the generalised Q-decomposition of \citet[Lemma~4]{tianpearl2003}. Let
$V_1<\dots<V_n$ be the vertices of the call under consideration, that is, the elements of its own
vertex set $\V$, which along the tail is a proper subset of the vertex set $\Vz$ of the query. The
lemma states that
\begin{equation}\label{eq:TP}
  Q(V_i\mid C_i)=\frac{Q[\pre{i}]}{Q[\pre{i-1}]},\qquad
  Q[\pre{i}]=\sum_{V_{i+1},\dots,V_n}Q,
\end{equation}
where $Q[\pre{0}]=1$ and the sets $\pre{i}$ are called the \emph{prefixes} of $\V$; in its item
(iii), the lemma adds that this ratio is a function of $V_i$ and $C_i$ only. As written, \eqref{eq:TP} divides by the
marginal on the whole prefix $\pre{i-1}$, although by item (iii) the factor depends only on $C_i$.
What matters is therefore the prefix that ends at the last vertex of $C_i$ in the order $\pi$, which
we denote by $V_\ell$, and \eqref{eq:TP} can be rewritten accordingly. Since the factor does not depend
on the vertices $V_{\ell+1},\dots,V_{i-1}$, the identity $Q[\pre{i}]=Q(V_i\mid C_i)\,Q[\pre{i-1}]$ can
be summed over them, which gives
\begin{equation}\label{eq:short}
  Q(V_i\mid C_i)=\frac{\sum_{V_{\ell+1},\dots,V_{i-1}}Q[\pre{i}]}{Q[\pre{\ell}]} .
\end{equation}
Two special cases are worth noting. When $C_i$ is empty there is no denominator, and the summation
runs over all of $\pre{i-1}$; when $V_{i-1}$ belongs to $C_i$ (as it does for every nonempty $C_i$
in the examples of this paper), $\ell=i-1$ and \eqref{eq:short} coincides with \eqref{eq:TP}. The
following example shows what these formulas give when $Q$ is a summed product.

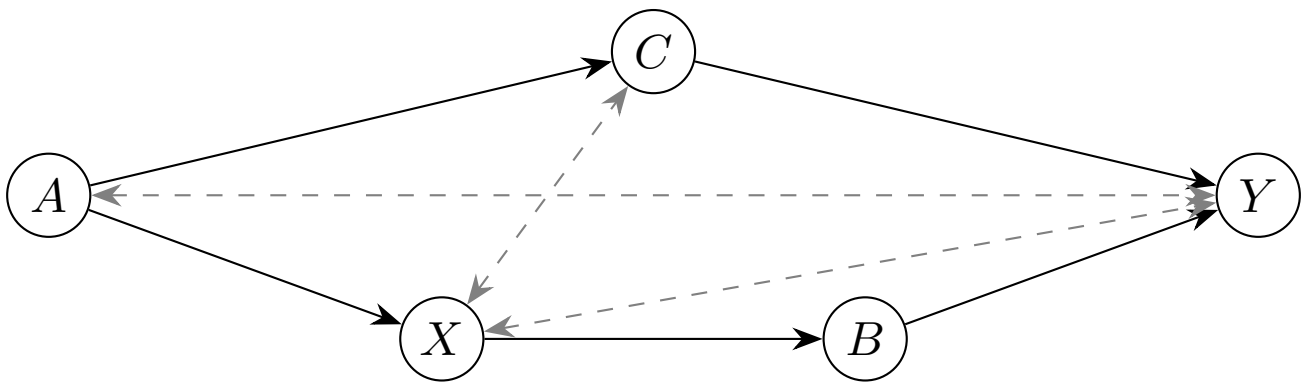
\begin{figure}[!ht]\centering
\begin{tikzpicture}[x=1cm,y=1cm]
\node[vtx] (A) at (0,0) {$A$};
\node[vtx] (C) at (4,0.95) {$C$};
\node[vtx] (X) at (2.6,-0.95) {$X$};
\node[vtx] (B) at (5.4,-0.95) {$B$};
\node[vtx] (Y) at (8,0) {$Y$};
\draw[dir] (A) -- (C);
\draw[dir] (C) -- (Y);
\draw[dir] (A) -- (X);
\draw[dir] (X) -- (B);
\draw[dir] (B) -- (Y);
\draw[bid] (A) -- (Y);
\draw[bid] (X) -- (C);
\draw[bid] (X) -- (Y);
\end{tikzpicture}
\caption{The graph of Example~\ref{ex:cancel}; dashed edges are bidirected.}
\label{fig:graphs}
\end{figure}

\begin{example}\label{ex:cancel}
Consider the query $P(Y\mid\hat A,\hat X,\hat C)$ on the graph of Figure~\ref{fig:graphs}. The
root executes line~4, since the graph without $A$, $X$ and $C$ has the c-components $\{B\}$ and
$\{Y\}$. The branch for $B$ has the trajectory $4,2,6$ and returns $P(B\mid X)$. The branch for $Y$
has the trajectory $4,7,2,7,2,1$. Its first line~7 finds the c-component $\Sp=\{A,X,C,Y\}$ of $Y$, and
$B$ becomes a fixed value. Inside this component the only directed edges are $A\to X$, $A\to C$ and
$C\to Y$, so that $X$ is not an ancestor of $Y$, and the line~2 that follows sums it out. The next
call has the vertex set $\V=\{A,C,Y\}$, with the edges $A\to C$, $C\to Y$ and $A\leftrightarrow Y$. It
executes line~7 with $\Sp=\{A,Y\}$, so that it has to form the Tian factors of $A$ and of $Y$, whose
conditioning sets are $C_1=\varnothing$ and $C_3=\{A,C\}$.

Take first the topological order $A<X<C<B<Y$. The first line~7 forms the terms $P(A)$, $P(X\mid A)$,
$P(C\mid A,X)$ and $P(Y\mid A,X,C,B)$, so that the distribution argument of the second line~7 is
\[
  Q=\sum_X P(A)\,P(X\mid A)\,P(C\mid A,X)\,P(Y\mid A,X,C,B).
\]
The vertices of $\V$ are $V_1=A$, $V_2=C$ and $V_3=Y$, and its prefixes are $\{A\}$, $\{A,C\}$ and
$\{A,C,Y\}$. By \eqref{eq:TP},
\begin{align*}
  Q[A,C,Y]&=Q,\\
  Q[A,C]&=\sum_Y Q=\sum_X P(A)\,P(X\mid A)\,P(C\mid A,X),\\
  Q[A]&=\sum_{C,Y} Q=P(A).
\end{align*}
In $Q[A,C]$ the term of $Y$, which no other term mentions, has summed to one, whereas the summation
over $X$ remains, because $X$ lies below $C$ and occurs in the term of $C$. In $Q[A]$ the terms of
$Y$, $C$ and $X$ have summed to one in turn. The Tian factors are therefore $Q(A)=P(A)$ and
\[
  Q(Y\mid A,C)=\frac{Q[A,C,Y]}{Q[A,C]}
  =\frac{\sum_X P(A)\,P(X\mid A)\,P(C\mid A,X)\,P(Y\mid A,X,C,B)}
        {\sum_X P(A)\,P(X\mid A)\,P(C\mid A,X)},
\]
and after the last line~2 has summed out $A$, the branch returns $\sum_A P(A)\,Q(Y\mid A,C)$, a
summation placed over a ratio. The same result is obtained in the topological order $A<X<B<C<Y$.

Take now the order $A<C<X<B<Y$. The first line~7 forms $P(A)$, $P(C\mid A)$, $P(X\mid A,C)$ and
$P(Y\mid A,C,X,B)$, and $Q$ is the sum over $X$ of their product. The set $\V$ and its prefixes are
the same, but $X$ now lies above $C$, so that in $Q[A,C]=\sum_YQ$ the terms of $Y$ and of $X$ sum to
one in turn, and $Q[A,C]=P(A)\,P(C\mid A)$. This product, which carries no summation, divides
$Q[A,C,Y]=P(A)\,P(C\mid A)\sum_X P(X\mid A,C)\,P(Y\mid A,C,X,B)$, so that
$Q(Y\mid A,C)=\sum_X P(X\mid A,C)\,P(Y\mid A,C,X,B)$, and ID returns
\begin{equation}\label{eq:cancelout}
  P(Y\mid\hat A,\hat X,\hat C)=\sum_B P(B\mid X)\sum_{A,X}P(A)\,P(X\mid A,C)\,P(Y\mid A,X,B,C),
\end{equation}
which contains no quotient. The two formulas are equal, and they differ in form only because the
summed vertex $X$ lies below $C$, the last vertex of $C_3$, in the first order and above it in the
other.
\end{example}

The example already displays the main features of the general case, namely that the factors formed
after a line~7 and a line~2 are ratios of sums of one and the same product, and that the position of
the summed vertices in the topological order decides whether the sums of the denominators disappear.
To describe the general case we first fix the notation for such a call.

\begin{definition}\label{def:round}
Let $c_1,\dots,c_t$ be the calls of a trajectory that execute line~6 or line~7, in their order. For
$s\ge2$ the call $c_s$ is a \emph{round}. By Theorem~\ref{thm:T}, $c_{s-1}$ executes line~7 and is
separated from $c_s$ by a single call, which executes line~2. The line~7 of $c_{s-1}$ retains a
c-component $\Sp$ and forms a factor $q_h$ for each of its vertices $h$. The line~2 that follows
splits $\Sp$ into the set $\Ub$ that it sums out and the set $\V$ that it keeps, so that the round
$c_s$ has the vertex set $\V=\Sp\setminus\Ub$ and the distribution argument
$Q=\sum_{\Ub}\prod_{h}q_h$. We write $V_1<\dots<V_n$ for the vertices of $\V$, and
$\Kb\subseteq\V$ for the set of the vertices whose Tian factors the round forms, which is its own
set $\Sp$ if it executes line~7 and $\Es$ if it executes line~6.
\end{definition}

The following scheme summarises where the sets of a round come from.
\begin{center}\small
\begin{tabular}{@{}ccccc@{}}
line 7 (call $c_{s-1}$) & $\longrightarrow$ & line 2 & $\longrightarrow$ & line 6 or 7 (the round $c_s$)\\[1pt]
retains $\Sp$, forms the $q_h$ & & sums out $\Ub$, keeps $\V=\Sp\setminus\Ub$ & & forms the factors of $\Kb\subseteq\V$
\end{tabular}
\end{center}
In Example~\ref{ex:cancel} the second line~7 is a round. The first line~7 retains
$\Sp=\{A,X,C,Y\}$, whose four terms are the factors $q_h$, line~2 sums out $\Ub=\{X\}$ and keeps
$\V=\{A,C,Y\}$, and the round forms the factors of $\Kb=\{A,Y\}$.

In a round, the sums \eqref{eq:TP} are computed as in the example, thanks to two properties of the
factors $q_h$. The distribution argument that a line~7 passes on is
always a joint distribution of the vertices that it retains, written as a product of conditionals
taken in the order $\pi$, so that each factor $q_h$ sums to one over its head and involves only
variables up to its head. The terms have these two properties, and \eqref{eq:TP} passes them on to the
factors of every round, because $Q[\pre{i}]$ involves only the variables of $\pre{i}$ and sums to
$Q[\pre{i-1}]$ over $V_i$. Summing $Q$ over the vertices of $\V$ above $V_i$ therefore removes, from
the highest head down, every factor whose head lies above $V_i$, and leaves\footnote{These
summations, over a variable that occurs only in a factor that sums to one over it, are trivial, and
here and below they are always carried out.}
\begin{equation}\label{eq:prefix}
  Q[\pre{i}]=\sum_{u\in\Ub,\ u<V_i}\ \prod_{h\le V_i}q_h .
\end{equation}
The sum on a prefix is thus a truncation of the product at $V_i$, which keeps the summations over the
vertices of $\Ub$ below $V_i$. In the first order of Example~\ref{ex:cancel}, the truncation at $C$
keeps the summation over $X$, and in the second it keeps no summation.

\subsection{Entangled rounds}

By \eqref{eq:short} and \eqref{eq:prefix}, the factor that a round forms for $V_i$ is the ratio of
two truncations of the same product, at $V_i$ and at the last vertex $V_\ell$ of $C_i$. Its
denominator is the shorter truncation, and it is a sum, rather than a plain product, exactly when a
vertex of $\Ub$ lies below $V_\ell$.

\begin{definition}\label{def:entangled}
A round is \emph{entangled at $V_i\in\Kb$} when $C_i$ is not empty and some vertex of $\Ub$ precedes
the last vertex of $C_i$ in the order $\pi$. A round is \emph{entangled} when it is so at one of its
vertices.
\end{definition}

Whether a round is entangled can be read off its last vertex alone. To see this, note that $V_n$ is an
outcome,
because every vertex of a round is an ancestor of $\Y$ in the graph of the round and no vertex lies
above $V_n$, and the c-component of an outcome in this graph is $\Kb$. The vertex $V_{n-1}$ then
belongs to $C_n$, because it is a member of $\Kb$, or a parent of $V_n$, or both; indeed, if it is not
a member of $\Kb$ it is not an outcome, and its only way to reach $\Y$ passes through $V_n$, the only
vertex above it. The last vertex of $C_n$ is therefore $V_{n-1}$ (so that, for $V_n$,
\eqref{eq:short} always has $\ell=n-1$ and coincides with \eqref{eq:TP}), and the last vertex of
every other set $C_i$ precedes it, so that
\begin{center}
\emph{a round is entangled if and only if some vertex of $\Ub$ precedes $V_{n-1}$,}
\end{center}
that is, if and only if a summed vertex has at least two vertices of $\V$ above it. In
Example~\ref{ex:cancel}, where $V_{n-1}=C$, the round is entangled in the two orders in which $X$
precedes $C$, and it is not entangled in the order $A<C<X<B<Y$. The proposition below gives the form
of the factors of a round in the absence of entanglement.

\begin{proposition}\label{thm:ent}
Consider a round, with the notation of Definition~\ref{def:round}.
\begin{enumerate}[label=(\roman*),nosep]
  \item If the round is not entangled, it creates no new ratio. The factor of every vertex of $\Kb$
    other than $V_n$ is the factor $q_h$ that the previous call formed for that vertex, unchanged,
    and the factor of $V_n$ absorbs the summation,
    \begin{equation}\label{eq:noquot}
      Q(V_n\mid C_n)=\sum_{\Ub'}\ q_{V_n}\prod_{u\in\Ub'}q_u ,
    \end{equation}
    where $\Ub'$ is the set of the vertices of $\Ub$ that precede $V_n$.
  \item In any round, the factor of a vertex $V_i$ at which the round is not entangled is formed
    without any division, and reads
    \[
      Q(V_i\mid C_i)=\sum_{V_\ell<h<V_i}\ \ \prod_{V_\ell<h\le V_i}q_h ,
    \]
    where $h$ runs over the vertices of $\Ub\cup\V$ and $V_\ell$ is the last vertex of $C_i$, the
    lower bounds being dropped when $C_i$ is empty.
\end{enumerate}
\end{proposition}

The proof, a direct computation with \eqref{eq:TP}, \eqref{eq:short} and \eqref{eq:prefix}, is given
in Appendix~\ref{app:proofs}. The factors $q_h$ in these formulas are those of the previous call, and
they may themselves be ratios when an earlier round was entangled; a round that is not entangled hands
such ratios on as they are and adds none of its own.

Two consequences of these formulas are useful in practice, the first of which concerns the factors
that remain ratios. Written as in \eqref{eq:TP}, the factors of two consecutive
vertices $V_{i-1}$ and $V_i$ of $\V$ that both belong to $\Kb$ share a sum, the denominator of the
second being the numerator of the first, so that in a product of the factors of consecutive vertices
of $\Kb$ only the last numerator and the first denominator remain (in graph~(b) of
Example~\ref{ex:tail} below, three ratios reduce in this way to one). An implementation should carry
out this cancellation, which involves equal sums and requires no test.

The second consequence concerns the trajectories none of whose rounds is entangled. Along such a
trajectory, statement (i) shows that each round hands on unchanged the terms formed by $c_1$, except for the factor of the last vertex, which absorbs
the summation over the vertices that have just been summed out, and the successive summations can be
merged into one. The part of the output produced along such a trajectory is then a single summation
over a product of terms $P(V_i\mid C_i)$ formed by $c_1$, and a query without entangled rounds is
answered by a formula that contains no quotient. The following example shows a tail of two rounds,
first without and then with entanglement, and makes both consequences explicit.

\begin{figure}[!ht]\centering
\newcommand{\pairgraph}[1]{%
\path[use as bounding box] (-0.4,-2.05) rectangle (6.9,1.6);
\foreach \i/\x in {1/0,2/2.2,3/4.4}{
  \node[vtx] (h\i) at (\x,1.5) {$H_\i$};
  \node[vtx] (p\i) at (\x,-1.5) {$Z_\i$};
  \draw[dir] (h\i) -- (p\i);
}
\node[vtx] (y) at (6,0) {$Y$};
\draw[dir] (p1) -- (y);
\draw[dir] (p2) -- (y);
\draw[dir] (p3) -- (y);
\draw[bid] (h1) -- (h2);
\draw[bid] (h2) -- (h3);
\draw[bid] (h1) -- (p2);
\draw[bid] (h2) -- (p3);
\draw[bid] (h1) -- (y);
\draw[bid] (h2) -- (y);
\draw[bid] (h3) -- (y);
\node[font=\small] at (3.25,-1.88) {(#1)};}
\begin{tikzpicture}[x=1cm,y=1cm]
\pairgraph{a}
\end{tikzpicture}\hspace{8mm}%
\begin{tikzpicture}[x=1cm,y=1cm]
\pairgraph{b}
\draw[dir] (p1) -- (h2);
\draw[dir] (p2) -- (h3);
\end{tikzpicture}
\caption{The two graphs of Example~\ref{ex:tail}; dashed edges are bidirected. Graph~(b) is
graph~(a) with the two additional edges $Z_1\to H_2$ and $Z_2\to H_3$.}
\label{fig:pairs}
\end{figure}
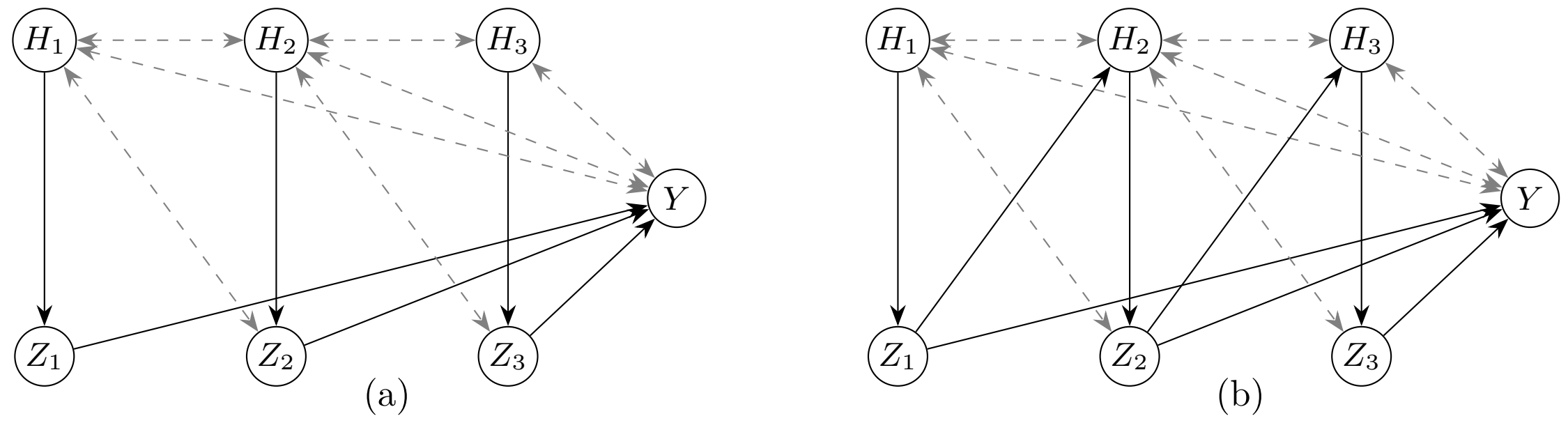

\begin{example}\label{ex:tail}
Consider the query $P(Y\mid\hat H_1,\hat Z_1,\hat H_2,\hat Z_2,\hat H_3,\hat Z_3)$ on each of the two
graphs of Figure~\ref{fig:pairs}, which give rise to the same trajectory. The root call has
$\Es=\Y=\{Y\}$ and satisfies (A) and (B), so that neither line~3 nor line~4 is executed, and the
trajectory is $7,2,7,2,7,2,1$, a tail of the maximal length $\lfloor6/2\rfloor$ allowed by
Theorem~\ref{thm:T}. The first line~7 retains the c-component $\{H_1,H_2,Z_2,H_3,Z_3,Y\}$ of $Y$, so
that $Z_1$ becomes a fixed value, and the line~2 that follows sums out $H_1$, which reaches $Y$ only
through $Z_1$. The second line~7 is a round with $\Ub=\{H_1\}$ and $\V=\{H_2,Z_2,H_3,Z_3,Y\}$. It
retains $\Kb=\{H_2,H_3,Z_3,Y\}$, so that $Z_2$ becomes a fixed value, and line~2 sums out $H_2$, which
reaches $Y$ only through $Z_2$. The third line~7 is a round with $\Ub=\{H_2\}$, $\V=\{H_3,Z_3,Y\}$
and $\Kb=\{H_3,Y\}$, after which line~2 sums out $H_3$ and line~1 returns. The two graphs differ in
the topological orders that they admit, and therefore in the formulas that ID returns.

\emph{Graph~(a).} Take the order $H_3<Z_3<H_2<Z_2<H_1<Z_1<Y$, in which the first line~7 forms the six
terms
\begin{gather*}
  P(H_3),\quad P(Z_3\mid H_3),\quad P(H_2\mid H_3,Z_3),\quad P(Z_2\mid H_2),\\
  P(H_1\mid H_3,Z_3,H_2,Z_2),\quad P(Y\mid H_3,Z_3,H_2,Z_2,H_1,Z_1).
\end{gather*}
In the first round $H_1$ lies above $Z_2$, the last but one vertex of $\V$, so that the round is not
entangled. By Proposition~\ref{thm:ent}(i) the factors of $H_3$, $Z_3$ and $H_2$ are their terms, and
the factor of $Y$ is, by \eqref{eq:noquot},
$\sum_{H_1}P(H_1\mid H_3,Z_3,H_2,Z_2)\,P(Y\mid H_3,Z_3,H_2,Z_2,H_1,Z_1)$. In the
second round $H_2$ lies above $Z_3$, so that this round is not entangled either, the factor of $H_3$
is again $P(H_3)$, and the factor of $Y$ is
\begin{align*}
  &\sum_{H_2}P(H_2\mid H_3,Z_3)\sum_{H_1}P(H_1\mid H_3,Z_3,H_2,Z_2)\,P(Y\mid H_3,Z_3,H_2,Z_2,H_1,Z_1)\\
  &\qquad=\sum_{H_1,H_2}P(H_2\mid H_3,Z_3)\,P(H_1\mid H_3,Z_3,H_2,Z_2)\,P(Y\mid H_3,Z_3,H_2,Z_2,H_1,Z_1),
\end{align*}
where the two summations merge because $H_1$, which lies above $H_2$, does not occur in the term of
$H_2$. The summation over $H_3$ merges with them for the same reason, and line~1 returns
\begin{equation}\label{eq:unent}
\begin{split}
  &P(Y\mid\hat H_1,\hat Z_1,\hat H_2,\hat Z_2,\hat H_3,\hat Z_3)\\
  &\qquad=\sum_{H_1,H_2,H_3}P(H_3)\,P(H_2\mid H_3,Z_3)\,P(H_1\mid H_3,Z_3,H_2,Z_2)\,
   P(Y\mid H_1,Z_1,H_2,Z_2,H_3,Z_3),
\end{split}
\end{equation}
a single summation over a product of four of the six terms, the terms of $Z_2$ and $Z_3$ having left
the product together with their heads.

\emph{Graph~(b).} The two additional edges leave a single topological order,
$H_1<Z_1<H_2<Z_2<H_3<Z_3<Y$. In it $H_1$ precedes every vertex of the first round and $H_2$ every
vertex of the second, so that both rounds are entangled. The first line~7 forms the six terms
\begin{gather*}
  f_1=P(H_1),\quad f_2=P(H_2\mid H_1,Z_1),\quad f_3=P(Z_2\mid H_1,Z_1,H_2),\quad
  f_4=P(H_3\mid H_1,Z_1,H_2,Z_2),\\
  f_5=P(Z_3\mid H_1,Z_1,H_2,Z_2,H_3),\quad f_6=P(Y\mid H_1,Z_1,H_2,Z_2,H_3,Z_3).
\end{gather*}
By \eqref{eq:prefix}, the sums of the first round on the prefixes ending at $H_2$, $Z_2$, $H_3$,
$Z_3$ and $Y$ are $A_2,\dots,A_6$, where $A_k=\sum_{H_1}f_1\cdots f_k$, and its factors are $A_2$ for
$H_2$, $A_4/A_3$ for $H_3$, $A_5/A_4$ for $Z_3$ and $A_6/A_5$ for $Y$. The last three telescope, as
observed above (each numerator cancels the denominator of the next factor), and their product is
$A_6/A_3$. The sums of the second round on the prefixes ending at $H_3$, $Z_3$ and $Y$ are then $B_1$, $B_2$ and $B_3$, where $B_j=\sum_{H_2}A_2A_{j+3}/A_3$, its
factors are $B_1$ for $H_3$ and $B_3/B_2$ for $Y$, and the output is
\begin{equation}\label{eq:nested}
  P(Y\mid\hat H_1,\hat Z_1,\hat H_2,\hat Z_2,\hat H_3,\hat Z_3)=\sum_{H_3}\frac{B_1\,B_3}{B_2},
  \qquad B_j=\sum_{H_2}\frac{A_2\,A_{j+3}}{A_3},
  \qquad A_k=\sum_{H_1}f_1\cdots f_k .
\end{equation}
Three summations are nested, and the ratios contain ratios, where \eqref{eq:unent} has a single
summation and no ratio. None of the sums $A_k$ is a marginal of $P$, because the product lacks the
term $P(Z_1\mid H_1)$ of the fixed value $Z_1$, which depends on $H_1$. Adding further pairs $H_k$,
$Z_k$, connected to their neighbours and to $Y$ as in graph~(b), gives for every $m$ the trajectory
$(7,2)^m,1$, whose $m-1$ rounds are all entangled, and an output with $m$ nested summations and $m-1$ nested
ratios.
\end{example}

\subsection{Entanglement from the graph}

Whether a round is entangled depends on the topological order $\pi$, while the sets $\Ub$ and $\V$
of Definition~\ref{def:round}, like the outcome set $\Y$ of the round and the graph $G_0$ of the
query, are the same in every order. As we have seen, a round is not entangled in $\pi$ exactly when every vertex of
$\V$ other than the last one precedes every vertex of $\Ub$. An order with this property may or may
not exist, and the following theorem decides the matter from $\Ub$, $\V$, $\Y$ and ancestry in $G_0$.

\begin{theorem}[Entanglement from the graph]\label{thm:graph}
Consider a round with sets $\Ub$ and $\V$ and with outcome set $\Y$, and let $z\in\Y$ be one of its
outcomes.
\begin{enumerate}[label=(\roman*),nosep]
  \item Some topological order of $G_0$ leaves the round unentangled, with $z$ as the last outcome in
    this order, if and only if neither $z$ nor any vertex of $\Ub$ is an ancestor in $G_0$ of a
    vertex of $\V$ other than $z$.
  \item A single topological order of $G_0$ leaves all the rounds of the trajectory unentangled, with
    $z$ as the last outcome in this order, if and only if the condition of (i) holds at every round.
\end{enumerate}
\end{theorem}

Since the last vertex of $\V$ is always the last of the outcomes, a round can be freed of its
entanglement by the choice of the order exactly when the condition of (i) holds for one of its
outcomes; with a single outcome $Y$ there is no choice, and the condition is that no vertex of $\Ub$
be an ancestor in $G_0$ of a vertex of $\V$ other than $Y$. In other words, the vertices that a round sums out must not be ancestors of the vertices that it
retains, the last outcome excepted. The proof, given in Appendix~\ref{app:proofs}, is constructive,
in that it builds the order whenever the condition holds, by listing first the vertices of $\V$ other
than $z$ together with all their ancestors in $G_0$, and then the remaining vertices, among which are
$z$ and the vertices of $\Ub$ (for several rounds, whose sets $\V$ are nested, the same is done from
the innermost set outwards). The criterion is local, since it involves only the two sets of the
round, which the trajectory provides, and it is evaluated before any formula is computed.

We now apply the criterion to the two examples, beginning with Example~\ref{ex:cancel}, where
$\Ub=\{X\}$ and $\V=\{A,C,Y\}$. There the vertex $X$ is an ancestor of $B$ and of $Y$ only, the
condition holds, and the order that lists $A$ and $C$ first is $A<C<X<B<Y$. In graph~(a) of
Example~\ref{ex:tail}, $H_1$ is an ancestor of $Z_1$ and $Y$ only and $H_2$ of $Z_2$ and $Y$ only, so
that the condition holds at both rounds with $z=Y$; the construction lists $H_3$ and $Z_3$, then $H_2$
and $Z_2$, then the remaining vertices, and gives the order used in the example. In graph~(b), $H_1$
is an ancestor of $H_2$, through $Z_1$, and $H_2$ is an ancestor of $H_3$, through $Z_2$, so that the
condition fails at both rounds. These two paths show how a failure arises when there is a single
outcome. A summed vertex $u$ is not an ancestor of the outcome inside the component retained by the
previous line~7, whereas the vertices of the round are. A directed path of $G_0$ from $u$ to one of
them therefore has to leave that component, and to pass through a vertex that a line~7 of the tail has
removed (as $Z_1$ and $Z_2$ are here).

The same example shows that the entanglement of a round has no consequence for the rounds that follow
it, since each of the two additional edges of graph~(b) governs one round. If only $Z_1\to H_2$ is
added to graph~(a), the first round is entangled in every order, because $H_1$ becomes an ancestor of
$H_2$, while the second is not entangled in the order $H_3<Z_3<H_1<Z_1<H_2<Z_2<Y$. If only
$Z_2\to H_3$ is added, it is the second round that is entangled in every order, while the first is not
entangled in the order $H_2<Z_2<H_3<Z_3<H_1<Z_1<Y$. The reason is that the tests of two successive
rounds concern different paths. At a round with sets $\Ub$ and $\V$, the test asks whether a directed
path of $G_0$ leads from $\Ub$ into $\V\setminus\{z\}$. The vertices summed out before the next round
belong to $\V\setminus\{z\}$ themselves, as do the vertices that the next round retains, so that the
next test asks about paths between two parts of $\V\setminus\{z\}$, which the first test neither
constrains nor involves. Every round of a trajectory therefore has to be tested separately, which
Theorem~\ref{thm:graph} allows for all of them at once, since the trajectory provides all their sets.

What an entangled round does hand on to the next is its ratios, which are among the factors $q_h$ of
that round. A round that is not entangled leaves them as they are (Proposition~\ref{thm:ent}(i)), and
an entangled one places them inside new ratios, as on graph~(b) of Example~\ref{ex:tail} in formula
\eqref{eq:nested}; in either case they remain in the output unless a later simplification removes
them.

A last remark concerns the queries for which line~4 is executed, and whose recursion tree therefore
has several trajectories. Theorem~\ref{thm:graph} applies to each trajectory separately, but ID uses
a single topological order for all of them, and the requirements of two trajectories may be
incompatible. This happens, for instance, when one branch sums out a vertex $A$ and retains a vertex
$B$, while another sums out $B$ and retains $A$, since the first then needs $B$ to precede $A$ and the
second needs the opposite. Each branch could then be freed of its entanglement by an order of its own,
but no single order does so for both, and one of them keeps an entangled round.

\section{Simplifying the ratios}
\label{sec:simp}

Theorem~\ref{thm:graph} decides whether the entanglement of a round can be avoided, but it takes into
account only the trivial summations, those of a factor that sums to one over its head. The summations
in the denominator of an entangled ratio may still be carried out by other means, through the
conditional independences among its terms, and this is found only by attempting the simplification
\citep{tikkakarvanen2017,companion}. In Example~\ref{ex:cancel}, in the order $A<X<C<B<Y$, they can,
since the terms of the denominator join into $P(A,X,C)$, and the denominator becomes
$\sum_XP(A,X,C)=P(A,C)$. This section shows that a denominator freed of its summations in this way
never has to be kept, and then gives two properties of the terms of an output on which a simplifier
can rely, and by which the correctness of its result can be checked.

By \eqref{eq:prefix}, a ratio of a round, or the product of the ratios of consecutive vertices of
$\Kb$, has the form
\begin{equation}\label{eq:ND}
  \frac{N}{D},\qquad N=\sum_{\Ab}F,\qquad D=\sum_{\Ab\cup\Bb}F,
\end{equation}
where $F$ is a product of factors, $\Ab$ is a set of summed vertices and $\Bb$ is the set of the heads
that are free in $N$ and summed in $D$. We consider simplifications of $D$ made of steps of two
kinds, which together cover the four rewritings (R1)--(R4) of \citet{companion}. A
\emph{rearrangement} replaces the product, or a part of it, by an expression that is equal to it for
all values of the variables, and leaves the summation untouched. It may reorder and regroup the
factors (R3), assemble terms into a joint distribution by the chain rule and by conditional
independences (the joins and insertions of \citealp{tikkakarvanen2017,companion}), factorise a joint
distribution back into terms, or cancel equal factors. A \emph{summation} eliminates a summed
variable $s$ from a joint distribution $P(\mathbf{H}\mid C)$ of which it is one of the heads,
replacing it by $P(\mathbf{H}\setminus\{s\}\mid C)$, when $s$ occurs in no other factor. This is the
marginalisation (R4), applied once the other factors have been taken out of the sum (R2), and it
reduces to (R1) when $s$ is the only head. The following theorem states that whatever these steps achieve on the denominator can be
carried over to the numerator, by replaying them there one by one.

\begin{theorem}[Replay]\label{thm:replay}
Let $N$ and $D$ be as in \eqref{eq:ND}, where $F$ is a product of terms, some of which may be
divisors. If rearrangements and summations remove all the summations of $D$, the ratio $N/D$ equals
a formula without quotient,
\[
  \frac{N}{D}=\sum_{\Ab}\ \prod_{e\in\Ab\cup\Bb}P(e\mid C_e),
\]
with one term, for a suitable set $C_e$, for each variable summed out of $D$. More generally, if
these steps bring $D$ to the form $\sum_{\Rb}F'$, where $F'$ is a product of terms, then
$N=\sum_{\Ab}F'\prod_{e\in\Eb}P(e\mid C_e)$ with $\Eb=(\Ab\cup\Bb)\setminus\Rb$, and every factor of
$F'$ that does not involve $\Rb$ cancels from $N/D$.
\end{theorem}

In the proof, given in Appendix~\ref{app:proofs}, the chain rule sets aside one term for each variable
that is summed out of $D$. To see the theorem at work,
take the ratio $Q(Y\mid A,C)$ of Example~\ref{ex:cancel} in the order $A<X<C<B<Y$, for which
$F=P(A)\,P(X\mid A)\,P(C\mid A,X)\,P(Y\mid A,X,C,B)$, $\Ab=\{X\}$ and $\Bb=\{Y\}$. The denominator
$D=\sum_{X,Y}F$ is simplified in three steps: a summation over $Y$ removes the term of $Y$, a
rearrangement joins the three remaining terms into $P(A,X,C)$ by the chain rule, and a summation over
$X$ leaves $D=P(A,C)$, without summation. The replay sets aside one term at each summation,
$P(Y\mid A,X,C,B)$ at the first and $P(X\mid A,C)$ at the second, and the theorem gives
\[
  \frac{N}{D}=\sum_X P(X\mid A,C)\,P(Y\mid A,X,C,B),
\]
which is the factor that ID forms directly in the order $A<C<X<B<Y$. In the term $P(X\mid A,C)$ the
condition $C$ follows the head $X$ in the order of the computation; in general the terms
$P(e\mid C_e)$ need be neither minimal nor conditioned only on preceding variables, which a
simplifier relying on either property has to take into account. The ratio of an entangled round thus
survives simplification only if its denominator retains a summation that none of these steps removes.

The Replay theorem tells a simplifier what to do with a ratio once its denominator has been
simplified. To find the simplifications in the first place, it helps to know which terms can occur in
an output, and the two properties announced above concern them, the first being about conditioning
sets. Each round writes its factors
through those of the previous call, so that every term produced within a single branch is one of the
terms $P(V_i\mid C_i)$ formed by its call $c_1$. Within a single branch every head therefore carries
a single conditioning set, and the terms of the denominator of a ratio are among those of its
numerator, so that cancelling factors within a ratio compares identical terms. Different branches of
line~4, on the contrary, form their own terms and may give different conditioning sets to the same
head, so that a cancellation between branches may not be available. Since the terms
$P(V_i\mid C_i)$ are minimal under faithfulness, conditions that can be dropped are created only by
simplifying steps, such as those of Theorem~\ref{thm:replay}, after which minimality has to be
restored \citep{companion}.

The second property is a count that every output satisfies, in the spirit of a dimensional check. For
each variable, count $+1$ for every term of which it is the head and $-1$ for every summation over it,
adding the counts of the factors of a product and subtracting that of a denominator from that of its
numerator; the result is called the \emph{weight} of the variable. A term behaves like a density in its head and a summation like an integration, so that
the weight records whether an expression is still a density in the variable. By \eqref{eq:TP} and
\eqref{eq:prefix}, the factor that ID forms for $V_i$ has weight $1$ in $V_i$ and $0$ in every other
variable, and every summation of ID lowers a weight from $1$ to $0$, so that
\begin{center}
\emph{in every ID output, each outcome of the query has weight $1$ and every other variable has
weight $0$.}
\end{center}
Let us count on the nested formula \eqref{eq:nested} of Example~\ref{ex:tail}. The intervened variable
$Z_1$ is never a head and has weight $0$. In every $A_k$, the variable $H_1$ is the head of $f_1$ and
is summed once, which gives $1-1=0$, while each of the other heads of $f_2,\dots,f_k$ counts $+1$. The
two tables below count the remaining variables, first in $B_3=\sum_{H_2}A_2A_6/A_3$ and then in the
output $\sum_{H_3}B_1B_3/B_2$.
\begin{center}
\renewcommand{\arraystretch}{1.0}
\begin{tabular}{@{}lrrrrr@{}}
\toprule
in $B_3$ & $H_2$ & $Z_2$ & $H_3$ & $Z_3$ & $Y$ \\
\midrule
$A_2$ & $+1$ & $0$ & $0$ & $0$ & $0$ \\
$A_6$ & $+1$ & $+1$ & $+1$ & $+1$ & $+1$ \\
$1/A_3$ & $-1$ & $-1$ & $0$ & $0$ & $0$ \\
$\sum_{H_2}$ & $-1$ & $0$ & $0$ & $0$ & $0$ \\
\midrule
weight & $0$ & $0$ & $1$ & $1$ & $1$ \\
\bottomrule
\end{tabular}
\hspace{14mm}
\begin{tabular}{@{}lrrr@{}}
\toprule
in the output & $H_3$ & $Z_3$ & $Y$ \\
\midrule
$B_1$ & $+1$ & $0$ & $0$ \\
$B_3$ & $+1$ & $+1$ & $+1$ \\
$1/B_2$ & $-1$ & $-1$ & $0$ \\
$\sum_{H_3}$ & $-1$ & $0$ & $0$ \\
\midrule
weight & $0$ & $0$ & $1$ \\
\bottomrule
\end{tabular}
\end{center}
The rows of $B_1$ and $B_2$ in the second table are obtained as in the first, with $A_4$ and $A_5$ in
the place of $A_6$. The variable $H_2$ is the head of three factors of $B_3$, two in the numerator and
one in the denominator, and is summed once, which gives $1+1-1-1=0$. The outcome $Y$ ends with weight
$1$ and all the intervened variables with weight $0$, although several of them occur as heads many
times. The rewritings used in simplification preserve the weights, so that this count (a necessary,
though not sufficient, condition for an expression to be a correct answer to the query) provides an
inexpensive check on any implementation.\footnote{The same count holds for the outputs of IDC, since
dividing by the sum over the outcomes leaves the outcomes with weight $1$ and the conditioning
variables with weight $0$.}

\section{Depth}
\label{sec:depth}

The trajectories also bound how deeply the summations and ratios of an output are nested. The
\emph{nesting depth} of an expression is defined by
\[
  d\bigl(P(V\mid C)\bigr)=0,\quad
  d(E_1E_2)=d(E_1/E_2)=\max\{d(E_1),d(E_2)\},\quad
  d\Bigl(\sum\nolimits_{\Mb} E\Bigr)=d(E)+1,
\]
a summation over the empty set being omitted, and it has to be read up to the rearrangements that the
formula allows. Summation indices are vertices of $G$ and are not renamed, so that a factor $E'$
standing beside $\sum_{\Mb}E$ can be moved inside the summation only if no variable of $\Mb$ occurs
in $E'$, not even as a fixed value. In \eqref{eq:frontdoor}, $P(Z\mid X)$ contains the intervened
variable $X$, which is summed in the neighbouring factor, and the formula has depth two.
By the depth of an ID output we therefore mean the least depth among the expressions obtained from it
by such moves, by merging consecutive summations and by trivial summations.

The obstruction that keeps the two summations of \eqref{eq:frontdoor} apart occurs only between the
branches of line~4. Within a single branch, what becomes of the summation created by a line~2
depends on the round that follows it. If this round is not entangled, the summation is absorbed by
the factor of the last vertex (Proposition~\ref{thm:ent}(i)) and merges with the summation that
later encloses this factor, as on graph~(a) of Example~\ref{ex:tail}, so that no level is added. If
the round is entangled, the summation enters the numerator and the denominator of ratios, out of
which it cannot be moved, and the depth increases by at most one.

\begin{theorem}[Depth of an ID output]\label{thm:depth}
Let $e$ be the largest number of entangled rounds on a trajectory of the query, and $m$ the largest
length of a tail. The output has depth at most $1+e$ if line~4 is not executed and at most $2+e$ if it
is, and in both cases at most $1+m$.
\end{theorem}

The proof is given in Appendix~\ref{app:proofs}, and it follows at once from the theorem that an
output of depth three or more requires $e\ge1$ and $m\ge2$, that is, an entangled round and a tail of
at least two repetitions, which in turn requires an intervention set $\X_0$ of at least four vertices
at the beginning of the tail (Theorem~\ref{thm:T}). The bounds $2+e$ and $1+m$ are attained by the
front-door formula \eqref{eq:frontdoor}, in which line~4 is executed, $e=0$ and $m=1$, and the bound
$1+e$ by graph~(b) of Example~\ref{ex:tail} and its extensions, in which line~4 is not executed and
the depth is $m$ with $e=m-1$. Graph~(a) of the same example, with the same trajectory and no entangled round, gives depth
one. In
Example~\ref{ex:cancel} the branch for $Y$ has depth one in the order $A<C<X<B<Y$ and depth two in
the order $A<X<C<B<Y$, where the summation over $A$ is placed over a ratio.

The number $e$ counts the entangled rounds of a trajectory wherever they occur along it, since, as we
saw in Section~\ref{sec:nest}, the entanglement of one round implies nothing about the others; with
only one of the two additional edges of graph~(b), for instance, the orders given there leave a single
entangled round, and the output has depth at most two. The bound also refers to the formula as ID
writes it,
and therefore overestimates what remains after simplification, because Theorem~\ref{thm:replay}
removes the quotients of every entangled round whose denominators are freed of their summations. The
rounds that really force quotients inside summations are thus those whose entanglement resists
simplification, and we have not proved a bound on the depth in terms of these rounds alone.

For an implementation, Proposition~\ref{prop:onesum}, Proposition~\ref{thm:ent} and
Theorem~\ref{thm:depth} determine the mathematical structure of the output, and hence the kind of
object that is needed to store it. After the trivial summations, the output reads
$\sum_{\El}\prod_kF_k$, with one factor $F_k$ for each branch of line~4, or a single factor if line~4
is not executed. Without entangled rounds, each $F_k$ is a single summation over a product of terms,
and two nested levels of summation, without any quotient, suffice. With entangled rounds, the object
has to admit quotients inside summations, each quotient being the ratio of two sums
\eqref{eq:prefix} of the same product, nested to a depth that grows with the number of entangled
rounds, and the recursive probability object of \citet{pedemonte2021} has this form. The expressions
of \citet{tikkakarvanen2017} (which nest products and summations and admit a quotient only between
two expressions, as a last operation) do not, so that neither the first formula of
Example~\ref{ex:cancel} nor the nested formula \eqref{eq:nested} fits their grammar.

\section{Conclusion}
\label{sec:concl}

We set out to understand the formula returned by the ID algorithm at its source, in order to know what
such a formula can contain before it is stored or simplified. The answer lies in the trajectories of
the algorithm, and the possible trajectories are few. Every trajectory consists of a clean-up, at most one
branching, a tail in which lines 7 and 2 alternate, and an exit (Theorem~\ref{thm:T}), and this
pattern, together with the sets on which each line acts, is obtained from the graph and the query by
graph operations, independently of the topological order. The pattern accounts for every summation of
the output: one outer summation, written by line~4 or, when line~4 is not executed, by the exit, runs
in either case over the same set, which is read off the original graph, and all the others are
created along the tail, each by a line~2 that marginalises the distribution rebuilt by the preceding
line~7 (Proposition~\ref{prop:onesum}).

The ratios are created along the tail as well, and here the topological order matters. From the second
repetition of the tail on, that is, at every round, each factor is the ratio of two sums of the same product, truncated at two vertices of
the round, and its denominator carries a summation only where the round is entangled, that is, where
a summed vertex precedes the last conditioning vertex of the factor (Proposition~\ref{thm:ent}). Whether
the order can be chosen so as to avoid the entanglement of a round is decided by a test of ancestry in
the original graph, applied to the two sets of that round (Theorem~\ref{thm:graph}); the test has to
be made at every round, because the entanglement of one round implies nothing about the next. When
the entanglement cannot be avoided, the
summations of a denominator may still be removed by simplification, and a denominator that has been
freed of its summations can always be cancelled (Theorem~\ref{thm:replay}). Finally, the nesting depth
of the output exceeds one, or two when line~4 is executed, by at most the number of entangled
rounds (Theorem~\ref{thm:depth}).

For an implementation these results fix what has to be represented and where simplification can act:
quotients inside summations are needed only for entangled rounds, cancellation within a branch
compares identical terms, and the weights of Section~\ref{sec:simp} give a test that every output has
to pass.

% Paragraph on future work, kept but commented out:
% Several questions are left for future work. We do not know whether the summations in the denominators
% of a round that is entangled in every topological order can ever be removed by simplification. When
% no order avoids every entanglement, it would be useful to know which order leaves the fewest entangled
% rounds, and hence the smallest depth. Finally, the three kinds of simplification recalled in the
% introduction have still to be combined into a single procedure that exploits the shape of the output
% described here.

\section*{Acknowledgements}

The graphs and queries presented as examples were found with the help of the AI model Claude
(Anthropic), after dozens of hours of discussion under the guidance of the authors. The authors are
supported by UKRI Future Leaders Fellowship Grant No. 10128920.

\appendix

\section{Proofs}
\label{app:proofs}

As in the table of Section~\ref{sec:traj}, unprimed symbols refer to the call that executes the line
under consideration and primed symbols to the call that this line makes, its child. Thus
$\Es'=\V'\setminus\X'$ is the live set of the child and $G'=G[\V']$ its graph.

\begin{proof}[Proof of Lemma~\ref{lem:AB}]
A directed path of $G_{\bar\X}$ that starts outside $\X$ contains no vertex of $\X$, because each of its
later vertices is entered by an edge of the path, and $G_{\bar\X}$ has no edge entering $\X$. The test
of line~3 fails when every vertex of $\Es$ is an ancestor of $\Y$ in $G_{\bar\X}$. If (A) holds, the
path that it provides for a vertex $v\in\Es$ lies in $\Es$, hence outside $\X$, so that it is also a
path of $G_{\bar\X}$ and makes $v$ an ancestor of $\Y$ there; the test then fails. If, on the other
hand, the test fails, every $v\in\Es$ has a directed path to $\Y$ in $G_{\bar\X}$. This path avoids
$\X$ by the remark above, so that it lies in $\Es$, and it is a path of $G$, since $G_{\bar\X}$ is $G$
with some edges deleted; this is (A). Both implications have been proved one vertex at a time, so
that the same proof, applied to a single live vertex, shows that it is an ancestor of $\Y$ in
$G_{\bar\X}$ if and only if it has a directed path to $\Y$ inside $\Es$. This fact is used again
below, in the proof of Lemma~\ref{lem:trap} (for line~3) and in that of
Proposition~\ref{prop:onesum}. As for (B), the set $\Es$ contains $\Y$ and is not
empty, so that $G[\Es]$ has a c-component, and line~4 tests whether it has more than one.
\end{proof}

\begin{proof}[Proof of Lemma~\ref{lem:trap}]
\emph{Line~2.} Line~2 deletes the vertices that are not ancestors of $\Y$. Under (A) every live
vertex is an ancestor of $\Y$, so that the deleted vertices are all intervened, and the child has the
same live set as the call and the same graph induced on it (in formulas, $\V'=\An(\Y)_G$ and
$\X'=\X\cap\V'$, so that $\Es'=\V'\setminus\X'=\Es\cap\V'=\Es$, because $\Es\subseteq\V'$). Conditions (A) and (B) are properties of
this graph, in which the paths required by (A) lie, and they therefore pass to the child; so does
saturation, since the outcome set is not changed either.

\emph{Line~7.} Line~7 deletes the vertices outside $\Sp$, which are all intervened, because
$\Es\subseteq\Sp$. As at line~2, the live set and the graph induced on it are unchanged, so that (A)
and (B), which hold at every call that reaches line~7, hold at the child, and saturation is preserved.

\emph{Line~4.} The $i$-th child has the graph $G$, the outcome set $\Y'=\Es_i$ and the intervention set
$\V\setminus\Es_i$, so that $\Es'=\Es_i=\Y'$ and the child is saturated. Condition (A) then holds
through paths of length zero, and (B) holds because the vertices of $\Es_i$, a c-component of
$G[\Es]$, are connected by bidirected paths inside $\Es_i$.

\emph{Line~3.} The child has the graph $G$, the outcome set $\Y$ and the intervention set $\X\cup\Ew$,
so that $\Es'=\Es\setminus\Ew$ consists of the vertices of $\Es$ that are ancestors of $\Y$ in
$G_{\bar\X}$. A directed path from $v\in\Es'$ to $\Y$ in $G_{\bar\X}$ avoids $\X$, by the remark that
opens the proof of Lemma~\ref{lem:AB}, and every vertex of this path reaches $\Y$ along the remainder
of the path, hence belongs to $\Es'$. Each vertex of $\Es'$ thus reaches $\Y$ along a path that lies in
$\Es'$, which is (A) for the child. Nothing can be said about (B), for two reasons. The test of line~3
precedes that of line~4, so that (B) need not hold at the call itself; and even when $G[\Es]$ has a
single c-component, $G[\Es']$ may have several, because the bidirected paths that connected the
vertices of $\Es'$ may have passed through the removed set $\Ew$. The child of a line~3 can therefore
execute line~4.
\end{proof}

\begin{proof}[Proof of Lemma~\ref{lem:flow}]
\ref{it:22} The child of a line~2 has the vertex set $\An(\Y)_G$, and in the graph induced on this set
every vertex is still an ancestor of $\Y$, since a directed path to $\Y$ consists of ancestors of $\Y$.
The test of line~2 therefore fails at the child.

\ref{it:1} Line~1 requires an empty intervention set. The root has a nonempty one by assumption, and so
has the child of a line~3, where $\X'\supseteq\Ew\neq\varnothing$, and of a line~4, where
$\X'=\V\setminus\Es_i\supseteq\X\neq\varnothing$. The same holds for the child of a line~7, whose
intervention set $\X'=\X\cap\Sp$ consists of the vertices of $\Sp$ outside $\Es$ and is never empty,
because line~7 is executed only when $\Es$ is a proper subset of $\Sp$. A call that executes line~1
can therefore only be the child of a line~2.

\ref{it:3} A call that satisfies (A) does not execute line~3 (Lemma~\ref{lem:AB}), and if it executes
line 2, 4 or 7 it passes (A) on to its children (Lemma~\ref{lem:trap}), so that (A), once it holds,
holds at all the descendants. Since line~3 gives (A) to its child, it is executed at most once on a
trajectory. Moreover a call that executes line~3 does not satisfy (A), so that it is either the root
or the child of a line~2, the only line whose child can fail to satisfy (A), since lines 3, 4 and 7
give (A) to theirs. In the second case the call that executes this line~2 does not satisfy (A) either, since line~2
would have passed it on, and by \ref{it:22} it is not itself the child of a line~2, so that it is the
root. Line~3 is therefore executed at step 1 or 2.

\ref{it:7} A call that executes line~7 satisfies (A) and (B), and so does its child
(Lemma~\ref{lem:trap}). At the child, the test of line~1 fails by \ref{it:1} and those of lines 3 and 4
fail by Lemma~\ref{lem:AB}, whereas the test of line~5 succeeds, because the vertex set $\Sp$ of the
child is a single c-component of its graph. Since the query is identifiable, line~5 is not executed, so
that the child is stopped by an earlier test, and only that of line~2 is left.

\emph{Line~4.} A call that executes line~4 gives (A) and (B) to its children. Among the lines that
make further calls, a call satisfying both conditions can execute only lines 2 and 7, which pass them
on, so that no descendant of a line~4 executes line~4. Suppose now that two calls executed line~4. If
one were a descendant of the other, this would contradict what has just been shown. Otherwise their
paths from the root would separate at some call, which, being a branching point, would execute line~4
and have both of them among its descendants, which is again a contradiction, so that line~4 is
executed at most once.
\end{proof}

\begin{proof}[Proof of Theorem~\ref{thm:T}]
Let the clean-up be the initial run of lines 2 and 3 of the trajectory. By
Lemma~\ref{lem:flow}\ref{it:3}, line~3 occurs in it at most once, as the first call or as the second
after a line~2, and by Lemma~\ref{lem:flow}\ref{it:22} line~2 does not occur twice in a row. Nor does
line~2 follow line~3, which changes neither the graph nor the outcome set, so that the test of line~2,
which has just failed, fails again. The clean-up is therefore one of $\varnothing$, $2$, $3$ and
$2\,3$, and the call that follows it executes line 1, 4, 6 or 7. Lines 1 and 6 end the trajectory,
whereas line~4 is executed at most once (Lemma~\ref{lem:flow}), and not below a line~7, all of whose
descendants satisfy (A) and (B); hence it occurs, if at all, immediately after the clean-up. Each of its children
satisfies (A) and (B) and has a nonempty intervention set, so that it executes line 2, 6 or 7, and a
line~2 is not followed by another.

From this point on every call satisfies (A) and (B), and the next line is 6, 7 or, after a line~2,
line~1. A line~7 is followed by a line~2 (Lemma~\ref{lem:flow}\ref{it:7}), after which the same three
possibilities recur, so that the pair $7\,2$ is repeated until the trajectory exits at line~6 or at
line~1; line~1 follows a line~2 by Lemma~\ref{lem:flow}\ref{it:1}, and neither exit follows a line~7.

For the bound on $m$, the line~7 of a repetition deletes the set $\V\setminus\Sp$, which is not empty,
since otherwise the test of line~5 would have succeeded, and is contained in $\X$, since
$\Es\subseteq\Sp$. Its line~2 deletes the vertices of $\Sp$ that are not ancestors of $\Y$ in
$G[\Sp]$, a set which is not empty, since the line was executed, and is contained in the intervention
set, since by (A) all the live vertices are ancestors of $\Y$. The $2m$ sets deleted along the tail are
disjoint subsets of $\X_0$, so that $2m\le|\X_0|$, with equality when they are single vertices that
exhaust $\X_0$, as in Example~\ref{ex:tail} ($|\X_0|=6$, $m=3$).
\end{proof}

\begin{proof}[Proof of Proposition~\ref{prop:onesum}]
The form of the output and the origin of its summations were established before the statement, and
only the expression of $\El$ remains to be proved. Let $c^{*}$ be the call at which the clean-up ends, that is, the first call after it, and
$\Es^{*}$ its live set. The outcome set of this call is that of the query, and the output is summed over its
leftover set, so that $\El=\Es^{*}\setminus\Y$. Let $\Db=\An(\Y)_{G_0[\Vz\setminus\X]}$ be the set of
the vertices from which $\Y$ can be reached by a directed path of $G_0$, the original graph, that
avoids the intervention set $\X$ of the query, the vertices of $\Y$ included. It suffices to prove
that $\Es^{*}=\Db$, and we do so by following along the clean-up the set $\Db(c)$ defined as follows.

For any call $c$, we say that a live vertex is \emph{internal} if it has a directed path to $\Y$
inside the live set, and \emph{external} otherwise, and we write $\Db(c)$ for the set of the internal
vertices of $c$. By the fact noted in the proof of Lemma~\ref{lem:AB}, the internal vertices are the
live vertices that are ancestors of $\Y$ in $G_{\bar\X}$, and condition (A) says that no live vertex
is external. At the root, $\Db(c)=\Db$ by the definition of $\Db$, since the live set of the root is
$\Vz\setminus\X$.

Neither line of the clean-up changes $\Db(c)$. Line~2 deletes only vertices that are not ancestors of
$\Y$, whereas an internal vertex, like every vertex of the path that makes it internal, is an ancestor
of $\Y$; the internal vertices and their paths thus survive, and no new one appears, since the child
has a smaller live set and a smaller graph. Line~3 keeps the graph and moves to the intervention set
exactly the external vertices, so that the live set of its child is $\Db(c)$; the vertices of a path
that makes a vertex internal are themselves internal, so that these paths survive as well, and the
child has the same internal vertices.

Finally, the call $c^{*}$ has no external vertex, as we check on the
four clean-ups of Theorem~\ref{thm:T}. If the clean-up ends with a line~3, this line has just moved
every external vertex to the intervention set. If the clean-up is empty, $c^{*}$ is the root, which does not execute line~1, because its
intervention set is not empty, nor line~2 or line~3; the test of line~3 is therefore reached and
fails, so that no live vertex is external (Lemma~\ref{lem:AB}). If the clean-up is a line~2 alone, $c^{*}$ is the child of this line~2, and the
same argument applies (the test of line~3 fails) unless its intervention set is empty; in that case $c^{*}$ executes line~1,
and all its vertices are live and are ancestors of $\Y$, hence internal. In every case the live set
$\Es^{*}$ of $c^{*}$ is $\Db(c^{*})$, which equals $\Db$ because $\Db(c)$ has not changed along the
clean-up.
\end{proof}

\begin{proof}[Proof of Proposition~\ref{thm:ent}]
(i) If the round is not entangled, every vertex of $\Ub$ lies above $V_{n-1}$. For $i<n$ the sum
\eqref{eq:prefix} then contains no vertex of $\Ub$ and reads $Q[\pre{i}]=q_{V_1}\cdots q_{V_i}$, so
that the ratio \eqref{eq:TP} is $q_{V_i}$. For $i=n$, the heads up to $V_n$ are $V_1,\dots,V_n$ and
the vertices of $\Ub'$, so that
\[
  Q[\pre{n}]=\sum_{\Ub'}\ q_{V_1}\cdots q_{V_{n-1}}\;q_{V_n}\prod_{u\in\Ub'}q_u .
\]
The factors $q_{V_1},\dots,q_{V_{n-1}}$ involve only variables up to $V_{n-1}$, hence no vertex of
$\Ub'$, and can be taken out of the summation. Their product is $Q[\pre{n-1}]>0$, and dividing by it leaves \eqref{eq:noquot}.

(ii) Suppose first that $C_i$ is not empty. Since the round is not entangled at $V_i$, no vertex of
$\Ub$ precedes $V_\ell$, so that the sum \eqref{eq:prefix} for the prefix $\pre{\ell}$ is the plain
product $\prod_{h\le V_\ell}q_h$, and this is the denominator of \eqref{eq:short}. In the numerator of
\eqref{eq:short}, $Q[\pre{i}]$ is given by \eqref{eq:prefix} as a sum over the vertices of $\Ub$
below $V_i$, and \eqref{eq:short} sums it further over $V_{\ell+1},\dots,V_{i-1}$; together these are
all the vertices $h$ of $\Ub\cup\V$ with $V_\ell<h<V_i$, since no vertex of $\Ub$ lies below
$V_\ell$. The factors $q_h$ with $h\le V_\ell$ involve only variables up to $V_\ell$, hence none of
the summed ones, so that their product can be taken out of the summations of the numerator, where it
cancels against the denominator, and what remains is the expression of the statement.

If $C_i$ is empty, \eqref{eq:short} has no denominator. Its numerator sums $Q[\pre{i}]$ over
$V_1,\dots,V_{i-1}$, and by \eqref{eq:prefix} $Q[\pre{i}]$ is itself the sum, over the vertices of
$\Ub$ below $V_i$, of the product of all the factors $q_h$ with $h\le V_i$. The two summations
together run over all the vertices $h<V_i$ of $\Ub\cup\V$, which gives the expression of the
statement without its lower bounds.
\end{proof}

\begin{proof}[Proof of Theorem~\ref{thm:graph}]
We write $\An(\mathbf{W})$ for the set of the vertices from which a directed path of $G_0$, possibly
of length zero, leads to a vertex of $\mathbf{W}$. The proof uses two facts. First, $\An(\mathbf{W})$
contains every parent of each of its vertices, that is, if $b\in\An(\mathbf{W})$ and $a\to b$ is an
edge of $G_0$, then $a\in\An(\mathbf{W})$. Second, as shown after Definition~\ref{def:entangled}, a
round is not entangled in an order $\pi$ if and only if every vertex of $\V$ other than the last one
precedes every vertex of $\Ub$ in $\pi$; moreover the last vertex of $\V$ is an outcome, and hence
the last of the outcomes, since $\V$ contains $\Y$.

(i) Suppose that the round is not entangled in a topological order $\pi$ in which $z$ is the last
outcome. By the second fact $z$ is the last vertex of $\V$, so that every vertex $v$ of
$\V\setminus\{z\}$ precedes $z$, and it precedes every vertex of $\Ub$. On the other hand, a directed
path of $G_0$ from a vertex $x$ to $v$ forces $x<v$, because every edge of $G_0$ goes from an earlier
to a later vertex of $\pi$. No such path can therefore start at $z$ or at a vertex of $\Ub$, which is
the condition of the statement.

Conversely, suppose that $z$ satisfies the condition, and let $\Ab=\An(\V\setminus\{z\})$. The
condition says precisely that $\Ab$ contains neither $z$ nor any vertex of $\Ub$. We build $\pi$ in
two blocks, listing first the vertices of $\Ab$ (which are thus the earliest vertices of $\pi$), in a
topological order of the subgraph that they induce, and then all the other vertices, among which are $z$ and the vertices of $\Ub$, in a
topological order of theirs. The result is a topological order of $G_0$, because an edge between two
vertices of the same block is respected by construction, and an edge $a\to b$ between the blocks
cannot lead from the second block to the first, since $b\in\Ab$ would imply $a\in\Ab$. In this order
every vertex of $\V\setminus\{z\}$, which lies in the first block, precedes $z$ and all the vertices
of $\Ub$, which lie in the second. Hence $z$ is the last vertex of $\V$, and therefore the last
outcome, the other vertices of $\V$ precede those of $\Ub$, and the round is not entangled.

With a single outcome $Y$, the vertex $z$ can only be $Y$. Every other vertex of $\V$ is an ancestor
of $Y$ in the graph of the round, hence in $G_0$, and a directed path from $Y$ to one of them would
close a directed cycle, so that the part of the condition that concerns $z$ holds automatically.

(ii) Let $\Ub_s$ and $\V_s$ be the sets of the rounds $c_s$, for $2\le s\le t$. The vertices of a
round are among those of the previous round, when there is one, so that
$\V_t\subseteq\dots\subseteq\V_2$, and every $\V_s$ contains $\Y$. If a single order leaves all the
rounds unentangled with $z$ as the last outcome, then $z$ is the last vertex of every $\V_s$, and the
first part of the proof of (i) gives the condition at every round.

Conversely, suppose that the condition holds at every round, and let
$\Ab_s=\An(\V_s\setminus\{z\})$, so that $\Ab_t\subseteq\dots\subseteq\Ab_2$. We now build $\pi$ in
$t$ blocks, listing first the vertices of $\Ab_t$ (the earliest of $\pi$), then those of $\Ab_{t-1}\setminus\Ab_t$, and so on
up to $\Ab_2\setminus\Ab_3$, and finally the vertices outside $\Ab_2$, each block in a topological
order of the subgraph that it induces. As before, an edge $a\to b$ cannot lead from a later block to
an earlier one, because $b$ then belongs to a set $\Ab_s$ that does not contain $a$, whereas $\Ab_s$
contains every parent of $b$. The result is therefore a topological order of $G_0$. For every $s$, the
vertices of $\V_s\setminus\{z\}$ lie in $\Ab_s$, whereas $z$ and the vertices of $\Ub_s$ lie outside
$\Ab_s$ by the condition, and are listed after all the vertices of $\Ab_s$. More precisely, $z$ lies
outside every $\Ab_s$, hence in the last block. The vertices of $\Ub_s$ lie in the block
$\Ab_{s-1}\setminus\Ab_s$ if $s\ge3$ (they are vertices of $\V_{s-1}$ other than $z$, hence belong to
$\Ab_{s-1}$, and they lie outside $\Ab_s$), and in the last block if $s=2$. Every round is therefore not entangled in this order, and $z$
is its last outcome. On graph~(a) of Example~\ref{ex:tail}, for instance, where $z=Y$ and $t=3$,
the blocks are $\Ab_3=\{H_3,Z_3\}$, $\Ab_2\setminus\Ab_3=\{H_2,Z_2\}$ and $\{H_1,Z_1,Y\}$, and
listing them in turn gives the order used there.
\end{proof}

\begin{proof}[Proof of Theorem~\ref{thm:replay}]
We prove the general statement, of which the first is the case $\Rb=\varnothing$, where $F'$ does not
involve $\Rb$ and cancels entirely. We follow the steps of the simplification of $D$, writing $F_k$
for its product after $k$ steps, and we maintain an identity $F=F_kL_k$, valid for all values of the
variables, in which $L_k$ is a product of one term $P(e\mid\cdot)$ for each variable $e$ summed out so
far, none of which occurs in $F_k$. Initially $F_0=F$ and $L_0=1$. A rearrangement replaces $F_k$ by an
expression $F_{k+1}$ equal to it for all values, and $L_k$ is kept. A summation replaces a joint
distribution $P(\mathbf{H}\mid C)$ of $F_k$, where $\mathbf{H}$ is the set of the heads joined in it,
by $P(\mathbf{H}\setminus\{s\}\mid C)$. By the chain rule, which
requires no condition,
$P(\mathbf{H}\mid C)=P(\mathbf{H}\setminus\{s\}\mid C)\,P(s\mid\mathbf{H}\setminus\{s\},C)$, so that
$F_k=F_{k+1}\,P(s\mid\mathbf{H}\setminus\{s\},C)$, and the last term is appended to $L_k$. After this
step $s$ no longer occurs in $F_{k+1}$, and later steps do not reintroduce it. At the end $F=F'L$ for
all values of the variables, where $L$ is the product of the terms set aside, one for each variable
of $\Eb$, whence $N=\sum_{\Ab}F'L$. A factor of $F'$ that does not involve $\Rb$ involves no variable
of $\Ab$ either, because the variables of $\Ab$ outside $\Rb$ have been summed out of $D$ and do not
occur in $F'$; such a factor can therefore be taken out of both summations and cancelled.
\end{proof}

\begin{proof}[Proof of Theorem~\ref{thm:depth}]
Fix a trajectory, and let $c_1,\dots,c_t$ be its calls that execute line~6 or line~7, as in
Definition~\ref{def:round}, so that $c_2,\dots,c_t$ are its rounds. We call \emph{sum over $c_s$} a
summation, possibly over the empty set, of a product of factors formed by $c_s$, and we write $D_s$ for the largest depth of such
an expression. The factors of $c_1$ are terms, of depth zero, and a summation of a product of terms
has depth one, so that $D_1=1$.

For $s\ge2$, by \eqref{eq:TP} and \eqref{eq:prefix} every factor of the round $c_s$ is a sum over
$c_{s-1}$ or the ratio of two of them, both of which have depth at most $D_{s-1}$, and a sum over
$c_s$ places one more summation on a product of such factors. A round thus adds at most one level,
that is, $D_s\le D_{s-1}+1$. If the round $c_s$ is not entangled, it adds none, and the bound improves
to $D_s\le D_{s-1}$. Indeed, by Proposition~\ref{thm:ent}(i) the factors of $c_s$ are those of
$c_{s-1}$, except for the factor \eqref{eq:noquot} of the last vertex $V_n$, which carries a
summation over $\Ub'$. The vertices of $\Ub'$ lie above every vertex of the round other than $V_n$,
and the factor of a vertex involves only variables up to that vertex, so that they do not occur in the
other factors of $c_s$. In a sum over $c_s$, the summation over $\Ub'$ can therefore be extended to
the whole product and merged with the outer summation, as on graph~(a) of Example~\ref{ex:tail}, and
the result is a sum over $c_{s-1}$. Since $D_1=1$, it follows that $D_s$ is at most one plus the
number of entangled rounds among $c_2,\dots,c_s$, so that $D_t\le1+e$, and that $D_s\le s$ for every
$s$.

Suppose first that line~4 is not executed. The output is then a sum over $c_t$, since the exit sums
the product of the factors of $c_t$ over $\El$ at line~6, and over $\El\cup\Ub$ at line~1, where
$\Ub$ is the set summed out by the line~2 that precedes it (if $t=0$, the exit must be line~1, which
returns a marginal of $P$, of depth zero). Its depth is therefore at most $D_t\le1+e$. This is also at
most $1+m$, because a tail of $m$ repetitions contains at most $m$ rounds, so that $e\le m$.

Suppose now that line~4 is executed. Each branch returns an expression of depth at most $1+e$, as we
have just seen, and line~4 sums the product of these expressions, which adds at most one level and
gives $2+e$. For the bound $1+m$ we use that the branches are saturated (Lemma~\ref{lem:trap}), so
that their exits sum over an empty leftover set. Consider a branch whose tail has $r\le m$
repetitions. If it exits at line~6, it returns the product of the factors of $c_{r+1}$, without any
summation, and each of these factors has depth at most $D_r$. If it exits at line~1, it returns a sum
over $c_r$, of depth at most $D_r$. In both cases the depth is at most $D_r\le r\le m$ (and zero if
$r=0$, where the branch returns terms or a marginal of $P$), and the summation of line~4 brings it to
at most $1+m$.
\end{proof}

\setlength{\bibsep}{3pt plus 1pt}

\end{document}